\documentclass[conference]{IEEEtran}

\DeclareMathAlphabet{\mathitbf}{OML}{cmm}{b}{it}

\usepackage{tikz}
\usetikzlibrary{matrix,arrows}

\usepackage{amsmath,amssymb,mathrsfs,amsthm}
\usepackage[colorlinks=true,linkcolor=black,anchorcolor=black,citecolor=black,filecolor=black,menucolor=black,urlcolor=black]{hyperref}
\usepackage{graphicx}
\usepackage{psfrag}
\usepackage{subfigure}
\usepackage{cite}
\usepackage{multirow}
\usepackage{stfloats}
\usepackage[margin=1in]{geometry}

\usepackage{epsfig}
\usepackage{amsfonts}
\usepackage{graphics}
\usepackage{color}
\usepackage{mathtools}
\DeclarePairedDelimiter{\ceil}{\lceil}{\rceil} 
\DeclarePairedDelimiter{\floor}{\lfloor}{\floor} 
\epsfverbosetrue

\usepackage{xspace}
\usepackage{bbm}
\catcode`@=11
\chardef\mathlig@atcode\count255

\def\actively#1#2{\begingroup\uccode`\~=`#2\relax\uppercase{\endgroup#1~}}
\def\mathlig@gobble{\afterassignment\mathlig@next@cmd\let\mathlig@next= }
\def\mathlig@delim{\mathlig@delim}
\def\mathlig@defcs#1{\expandafter\def\csname#1\endcsname}
\def\mathlig@let@cs#1#2{\expandafter\let\expandafter#1\csname#2\endcsname}
\def\mathlig@appendcs#1#2{\expandafter\edef\csname#1\endcsname{\csname#1\endcsname#2}}

\def\mathlig#1#2{\mathlig@checklig#1\mathlig@end\mathlig@defcs{mathlig@back@#1}{#2}\ignorespaces}

\def\mathlig@checklig#1#2\mathlig@end{%
 \expandafter\ifx\csname mathlig@forw@#1\endcsname\relax
 \expandafter\mathchardef\csname mathlig@back@#1\endcsname=\mathcode`#1%
 \mathcode`#1"8000\actively\def#1{\csname mathlig@look@#1\endcsname}%
 \mathlig@dolig#1\mathlig@delim
\fi
\mathlig@checksuffix#1#2\mathlig@end
}

\def\mathlig@checksuffix#1#2\mathlig@end{%
\ifx\mathlig@delim#2\mathlig@delim\relax\else\mathlig@checksuffix@{#1}#2\mathlig@end\fi
}
\def\mathlig@checksuffix@#1#2#3\mathlig@end{%
\expandafter\ifx\csname mathlig@forw@#1#2\endcsname\relax\mathlig@dosuffix{#1}{#2}\fi
\mathlig@checksuffix{#1#2}#3\mathlig@end
}

\def\mathlig@dosuffix#1#2{%
\mathlig@appendcs{mathlig@toks@#1}{#2}%
\mathlig@dolig{#1}{#2}\mathlig@delim
}

\def\mathlig@dolig#1#2\mathlig@delim{%
 \mathlig@defcs{mathlig@look@#1#2}{%
 \mathlig@let@cs\mathlig@next{mathlig@forw@#1#2}\futurelet\mathlig@next@tok\mathlig@next}%
 \mathlig@defcs{mathlig@forw@#1#2}{%
  \mathlig@let@cs\mathlig@next{mathlig@back@#1#2}%
  \mathlig@let@cs\checker{mathlig@chck@#1#2}%
  \mathlig@let@cs\mathligtoks{mathlig@toks@#1#2}%
  \expandafter\ifx\expandafter\mathlig@delim\mathligtoks\mathlig@delim\relax\else
  \expandafter\checker\mathligtoks\mathlig@delim\fi
  \mathlig@next
 }%
 \mathlig@defcs{mathlig@toks@#1#2}{}%
 \mathlig@defcs{mathlig@chck@#1#2}##1##2\mathlig@delim{%
  \ifx\mathlig@next@tok##1%
   \mathlig@let@cs\mathlig@next@cmd{mathlig@look@#1#2##1}\let\mathlig@next\mathlig@gobble
  \fi 
  \ifx\mathlig@delim##2\mathlig@delim\relax\else
   \csname mathlig@chck@#1#2\endcsname##2\mathlig@delim
  \fi
 }%
 \ifx\mathlig@delim#2\mathlig@delim\else
  \mathlig@defcs{mathlig@back@#1#2}{\csname mathlig@back@#1\endcsname #2}%
 \fi
}%

\catcode`@\mathlig@atcode

\newcommand{\muspace}{\mspace{1mu}}

\DeclareRobustCommand{\scond}{\mathchoice{\muspace\vert\muspace}{\vert}{\vert}{\vert}}
\mathlig{|}{\scond}

\DeclareRobustCommand{\discint}{\mathchoice{\mspace{-1.5mu}:\mspace{-1.5mu}}{\mspace{-1.5mu}:\mspace{-1.5mu}}{:}{:}}
\mathlig{::}{\discint}
\newcommand{\suchthat}{\mathchoice{\colon}{\colon}{:\mspace{1mu}}{:}}

\newcommand{\Gc}{\mathcal{G}}

\newcommand{\Ic}{\mathcal{I}}

\newcommand{\Cr}{\mathscr{C}}

\newcommand{\Rv}{{\bf R}}

\newcommand{\tv}{{\bf t}}
\newcommand{\xv}{{\bf x}}

\newcommand{\zv}{{\bf z}}

\newcommand{\xt}{{\tilde{x}}}

\def\a{\alpha}

\def\G{\Gamma}

\def\e{\epsilon}

\def\textiid{i.i.d.\@\xspace}
\newcommand\iid{\ifmmode\text{ i.i.d. } \else \textiid \fi}

\def\mathllap{\mathpalette\mathllapinternal}
\def\mathllapinternal#1#2{%
  \llap{$\mathsurround=0pt#1{#2}$}}

\def\clap#1{\hbox to 0pt{\hss#1\hss}}
\def\mathclap{\mathpalette\mathclapinternal}
\def\mathclapinternal#1#2{%
  \clap{$\mathsurround=0pt#1{#2}$}}

\let\oldstackrel\stackrel
\renewcommand{\stackrel}[2]{\oldstackrel{\mathclap{#1}}{#2}}

\renewcommand{\hbar}{h\mathllap{\overline{\vphantom{h}\hphantom{\rule{4.6pt}{0pt}}}\mspace{0.77mu}}}

\catcode`~=11 % make LaTeX treat tilde (~) like a normal character
\newcommand{\urltilde}{\kern -.06em\lower -.06em\hbox{~}\kern .02em}
\catcode`~=13 % revert back to treating tilde (~) as an active character

\newcommand{\lexprod}{\mathbin{\bullet}}

\def\0{\bf{0}}

\newtheorem{theorem}{Theorem}
\newtheorem{lemma}{Lemma}
\newtheorem{proposition}{Proposition}

\theoremstyle{definition}

\begin{document}

\title{Structural Properties of Index Coding Capacity Using Fractional Graph Theory}
\author{
\authorblockN{Fatemeh Arbabjolfaei and Young-Han Kim}
\authorblockA{Department of Electrical and Computer Engineering\\
University of California, San Diego\\
Email: \{farbabjo, yhk\}@ucsd.edu
}
}
\date{}
\maketitle

\begin{abstract}
The capacity region of the index coding problem is characterized through
the notion of confusion graph and its fractional chromatic number.
Based on this multiletter characterization, 
several structural properties of the capacity region are established,
some of which are already noted by Tahmasbi, Shahrasbi, and Gohari, 
but proved here with simple and more direct graph-theoretic arguments.
In particular, the capacity region of a given index coding problem
is shown to be simple functionals of the capacity regions of smaller
subproblems when the interaction between the subproblems is none,
one-way, or complete.
\end{abstract}

%%%%%%%%%%%%%%%%%%%%%%%%%%%%%%%%%%%%%%%%%%%%%%%%%%%%%%%%%%%%%%%%%%%%%%%%%%%%%%%%%%%%%%%%%%%%%%%%%
\section{Introduction}

Suppose that a sender wishes to communicate a tuple of $n$ messages, $x^n = (x_1, \ldots, x_n)$, $x_j \in \{0,1\}^{t_j}$, to their corresponding receivers using a shared noiseless channel.
Receiver $j \in [1:n] := \{1,2,\ldots, n\}$ has prior knowledge of a subset  $x(A_j) := \{x_i \suchthat i \in A_j\}$,  $A_j \subseteq [1:n] \setminus \{j\}$, of the messages and wishes to recover $x_j$.
It is assumed that the sender is aware of $A_1, \ldots, A_n$. 
The goal is to minimize the amount of information that should be broadcast from the sender to the receivers so that every receiver can recover its desired message.

\begin{figure}
\vspace{0.75em}
\begin{center}
\subfigure[]{
\small
\psfrag{1}[cb]{1}
\psfrag{2}[rc]{2}
\psfrag{3}[lc]{3}
\psfrag{4}{4}
\includegraphics[scale=0.45]{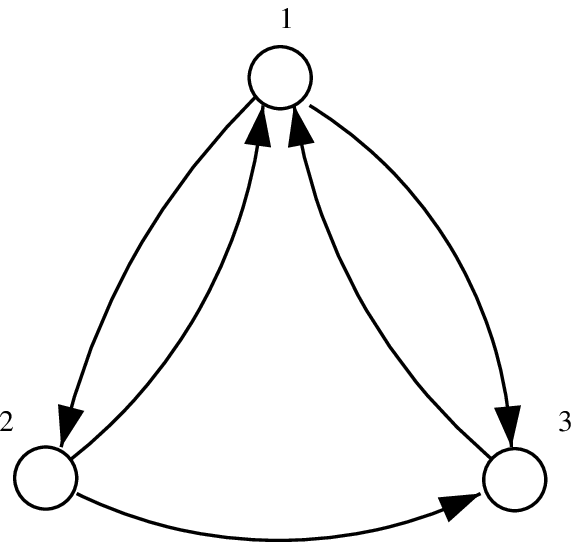}
}

\subfigure[]{
\vspace{0.5em}
\small
\psfrag{000}[cl]{000}
\psfrag{001}[cl]{001}
\psfrag{010}[l]{010}
\psfrag{011}[l]{011}
\psfrag{100}[clb]{100}
\psfrag{101}[clb]{101}
\psfrag{110}[]{110}
\psfrag{111}[]{111}
\includegraphics[scale=0.55]{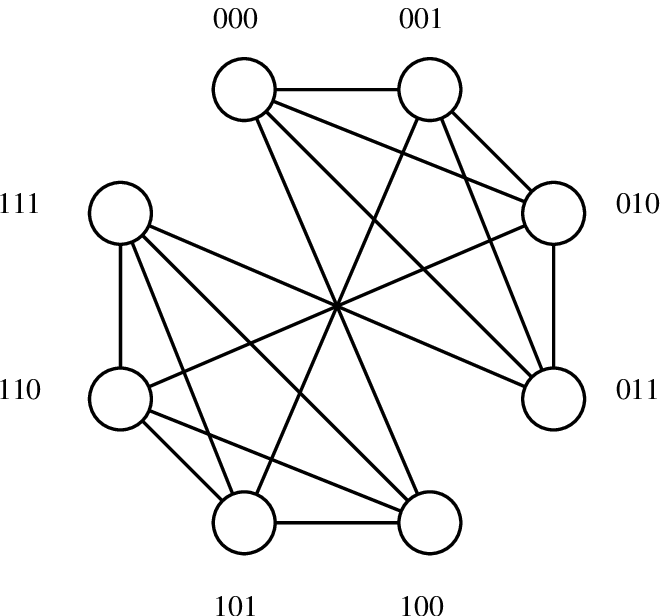}
}
\vspace{-1em}
\end{center}
\caption{(a) The  graph representation for the index coding problem with $A_1 = \{2,3\}, A_2 = \{1\}$, and $A_3 = \{1,2\}$. (b) The confusion graph corresponding to the integer tuple $(t_1, t_2, t_3) = (1,1,1)$. Each node is labeled with the a message tuple.}
\label{fig:3-message}
\vspace{-1em}
\end{figure}

Any instance of this problem, referred to collectively as the \emph{index coding problem}, 
is fully specified by the side information sets  $A_1, \ldots, A_n$. Equivalently, it can be specified
by a side information graph $G$ with $n$ nodes, in which
a directed edge $i \to j$ represents that receiver $j$ has message $i$ as side information, i.e., $i \in A_j$
(see Fig.~\ref{fig:3-message}(a)).
Thus, we often identify an index coding problem with its side information graph and simply write ``index coding problem $G$.''

A $(t_1, \ldots, t_n, r)$ index code is defined by
\begin{itemize}
\item an encoder $\phi: \prod_{i=1}^n \{0,1\}^{t_i} \to \{0,1\}^r$ that maps $n$-tuple of messages $x^n$
to an $r$-bit index and
\item $n$ decoders $\psi_j: \{0,1\}^r \times \prod_{k \in A_j} \{0,1\}^{t_k} \to \{0,1\}^{t_j}$ that maps the received index $\phi(x^n)$ and the side information $x(A_j)$ back to $x_j$ for $j \in [1::n]$.
\end{itemize}
Thus, for every $x^n \in \prod_{i=1}^n \{0,1\}^{t_i}$,
\[
\psi_j(\phi(x^n), x(A_j)) = x_j, \quad j \in [1::n].
\]
A rate tuple $(R_1,\ldots,R_n)$ is said to be \emph{achievable} for the index coding problem $G$
if there exists a $(t_1, \ldots, t_n, r)$ index code such that 
\[
R_j \leq \frac{t_j}{r}, \quad j \in [1:n].
\]
The \emph{capacity region} $\Cr$ 
of the index coding problem is defined as the closure of the set of achievable rate tuples.

Since Birk and Kol \cite{Birk--Kol1998} introduced the index coding problem in 1998, this simple yet fundamental
problem attracted several research communities (see \cite{Blasiak--Kleinberg--Lubetzky2013, Shanmugam--Dimakis--Langberg2013b, Maleki--Cadambe--Jafar2012a, Bar-Yossef--Birk--Jayram--Kol2011}
for a subset of recent contributions). The capacity region has been established for all 9,608 index coding problems
of $n = 5$ messages \cite{Arbabjolfaei--Bandemer--Kim--Sasoglu--Wang2013} (which includes all index coding problems \emph{up to} five messages by taking projections). However, the coding schemes developed for small $n$ prove to be
suboptimal when $n$ becomes large and there is no known computable characterization for the capacity region of a general index coding problem. On the theoretical side, there is no known
algorithm even to approximate the capacity region within a factor of $O(n^{1-\e})$ for $\e \in (0,1)$. 
On the computational side, the number of index coding problems blows up quickly in $n$ (for example,
there are 1,540,944 distinct instances of index coding problems for $n = 6$) and it also becomes quite challenging to
compare existing inner and outer bounds on the capacity region of each problem as $n$ increases.

As an intermediate step towards characterizing the capacity region (analytically, approximately, 
or numerically), we study some structural properties of the capacity region. 
In particular, we show that the side information graph $G$
can be partitioned into two vertex-induced subgraphs $G_1$ and $G_2$, then the capacity region $\Cr$ of the index coding problem $G$ can be characterized as a simple functional of the capacity regions $\Cr_1$ and $\Cr_2$ of
$G_1$ and $G_2$, respectively, provided that
\begin{enumerate}
\item there is no edge between $G_1$ and $G_2$, or
\item more generally, there is no edge from $G_2$ to $G_1$, or 
\item every node in $G_1$ is connected to every node in $G_2$ and vice versa.
\end{enumerate}
The immediate utility of these structural properties is that one can reduce the number of index coding problems
that need to be studied. For example, we can check (details not shown) that 1,366,783 (89\%) out of 
1,540,944 index coding problems for $n = 6$ fall into one of the three aforementioned criteria or another simple case (Proposition \ref{prop:ext-side-info} in Section \ref{sec:main}), 
significantly narrowing the set of problems that are worth further investigation.

We must note that the first two properties have been already established by Tahmasbi, Shahrasbi, and Gohari
\cite[Th.~2]{Tahmasbi--Shahrasbi--Gohari2014b} using a somewhat convoluted argument based on joint typicality encoding and covering. In comparison, our approach is more direct and based on the definition of the
capacity region itself. As discussed more precisely in Section \ref{sec:main}, our starting point is a 
graph-theoretic characterization of the index coding capacity region using the notion
of \emph{confusion graph}. This notion was introduced by 
Alon, Hassidim, Lubetzky, Stav, and Weinstein \cite{Alon--Hassidim--Lubetzky--Stav--Weinstein2008},
who characterized the optimal broadcast rate (the reciprocal of the symmetric capacity) using the chromatic
number of the confusion graph. We generalize and tighten their approach by connecting the capacity
region with the \emph{fractional} chromatic number of the confusion graph. This allows us to utilize
well-known results from fractional graph theory~\cite{Scheinerman--Ullman2011} such as the identities
on fractional chromatic numbers for graph products (see Section~\ref{sec:prelim}) to establish
several structural properties of the capacity region.
Our approach based on confusion graph and fractional chromatic number seems to be broadly applicable 
beyond these structural results. Although it is not presented here,
a similar method generalizes and tightens the recent result by Mazumdar~\cite{Mazumdar2014} on the
duality between index coding and distributed storage.

Throughout the paper, the base of logarithm is 2.

\section{Mathematical Preliminaries}
\label{sec:prelim}

\subsection{Confusion Graphs}

We generalize the notion of confusion graph,
which was originally introduced in \cite{Alon--Hassidim--Lubetzky--Stav--Weinstein2008} for equal-length messages.

Given an index coding problem $G$, 
two tuples of $n$ messages $x^n, z^n \in \prod_{i=1}^n \{0,1\}^{t_i}$ are said to be \emph{confusable
at receiver $j \in [1:n]$} if $x_j \ne z_j$ and $x_i = z_i$ for all $i \in A_j$.
We simply say $x^n$ and $z^n$ are \emph{confusable} if they are confusable at some receiver~$j$.
Given an index coding problem $G$ and 
a tuple of message lengths $\tv = (t_1, \ldots, t_n)$, the \emph{confusion graph}
$\G_{\tv}(G)$ is an undirected graph with $\prod_{i=1}^n 2^{t_i}$ vertices such that 
every vertex corresponds to a message tuple $x^n$ and 
two vertices are connected iff (if and only if) the corresponding 
message tuples are confusable. 

The confusion graph of the index coding problem with side information graph in Fig.~\ref{fig:3-message}(a) corresponding to $(t_1, t_2, t_3) = (1,1,1)$ is depicted in Fig.~\ref{fig:3-message}(b).

\subsection{Graph Coloring}

A (vertex) coloring of an undirected graph $\G$ 
is a mapping that assigns a color to each vertex
such that no two adjacent vertices share the same color.
The \emph{chromatic number} $\chi(\G)$ is the minimum number of colors such that a coloring of the graph exists.

More generally, a $b$-fold coloring assigns a set of $b$ colors to each vertex 
such that no two adjacent vertices share the same color.
The $b$-fold chromatic number $\chi^{(b)}(\G)$ is the minimum number of colors such that a $b$-fold coloring
exists.
The \emph{fractional chromatic number} of the graph is defined as
\[
\chi_f(\G) = \lim_{b \rightarrow \infty} \frac{\chi^{(b)}(\G)}{b} = \inf_b \frac{\chi^{(b)}(\G)}{b},
\]
where the limit exists since $\chi^{(b)}(\G)$ is subadditive.
Consequently,
\begin{equation} \label{eq:fractional}
\chi_f(\G) \le \chi(\G).
\end{equation}
Let $\Ic$ be the collection of all independent sets in $\G$ (i.e., sets of vertices such that
no two vertices are adjacent). 
The chromatic number and the fractional chromatic number 
are also characterized as the solution to the following optimization problem
\begin{equation*} 
\begin{split}
\text{minimize~}& \sum_{S \in \Ic} \rho_S\\
\text{subject to~} 
& \sum_{S \in \Ic \suchthat j \in S} \rho_S \ge 1, \quad j \in [1:n].
\end{split}
\end{equation*}
When the optimization variables $\rho_S$, $S \in \Ic$, take
integer values $\{0,1\}$, then the (integral) solution is the chromatic number. If this constraint is relaxed
and $\rho_S \in [0,1]$, then the (rational) solution is the fractional chromatic number \cite{Scheinerman--Ullman2011}.

\subsection{Graph Products}

Generally speaking, 
a graph product is a binary operation on two (undirected)
graphs $\G_1$ and $\G_2$ that produces a graph $\G$ on
the Cartesian product of the original vertex sets with the edge set constructed from
the original edge sets according to certain rules.
In this section, we review a few definitions of graph products and their (fractional) chromatic
numbers.
In the following, $v_1 \sim v_2$ denotes that there exists an edge between $v_1$ and $v_2$.
The notation $V(\G)$ means the vertex set of a graph $\G$.

The \emph{disjunctive product} $\G = \G_1*\G_2$ 
is defined as
$V(\G) = V(\G_1) \times V(\G_2)$ and $(u_1,u_2) \sim (v_1,v_2)$ iff
\[
u_1 \sim v_1 \quad \text{or} \quad u_2 \sim v_2.
\]
The fractional chromatic number of the disjunctive product is multiplicative.

\begin{lemma}[Scheinerman and Ullman {\cite[Cor.~3.4.2]{Scheinerman--Ullman2011}}]
\label{lem:orcoloring}
\[ 
\chi_f(\G_1*\G_2) = \chi_f(\G_1)\chi_f(\G_2). 
\]
\end{lemma}

Note that the chromatic number satisfies the following relationship~ \cite[Prop.~3.4.4]{Scheinerman--Ullman2011}:
\begin{equation} \label{eq:orcoloring2}
%\chi_f(\G_1)\chi(\G_2) \leq \chi(\G_1 * \G_2) \le \chi(\G_1)\chi(\G_2). 
\chi(\G_1 * \G_2) \le \chi(\G_1)\chi(\G_2). 
\end{equation}
The chromatic and fractional chromatic numbers of the power of a graph scale in the same exponential rate.

\begin{lemma}[Scheinerman and Ullman {\cite[Cor.~3.4.3]{Scheinerman--Ullman2011}}]
\label{lem:orcoloring1}
Let $\G^k$ be the $k$-th power of $\G$ in disjunctive product. Then
\[ 
\chi_f(\G) = \lim_{k \rightarrow \infty} \sqrt[k]{\chi(\G^k)} = \inf_{k} \sqrt[k]{\chi(\G^k)}. 
\]
\end{lemma}

The \emph{lexicographic product} $\G = \G_1 \lexprod \G_2$ is defined as 
$V(\G) = V(\G_1) \times V(\G_2)$ and $(u_1,u_2) \sim (v_1,v_2)$ iff
\[
u_1 \sim v_1 \quad \text{or} \quad (u_1 = v_1 \text{ and } u_2 \sim v_2).
\]
Note that the lexicographic product of graphs is not commutative. Nonetheless,
its fractional chromatic number is still multiplicative.

\begin{lemma}[Scheinerman and Ullman {\cite[Cor.~3.4.5]{Scheinerman--Ullman2011}}]
\label{lem:lexcoloring}
\[ 
\chi_f(\G_1 \lexprod \G_2) = \chi_f(\G_1)\chi_f(\G_2). 
\]
\end{lemma}

Note that the chromatic number satisfies the following relationship~ \cite[Th.~1]{Linial--Vazirani1989}:
\begin{equation} \label{eq:lexcoloring2}
\chi(\G_1 \lexprod \G_2) \le \chi(\G_1)\chi(\G_2). 
\end{equation}

The \emph{Cartesian product} $G = G_1 \wedge G_2$ is defined as
$V(\G) = V(\G_1) \times V(\G_2)$ and $(u_1,u_2) \sim (v_1,v_2)$ iff
\[
(u_1 = v_1 \text{ and }  u_2 \sim v_2) 
\quad \text{or} \quad (u_2 = v_2 \text{ and } u_1 \sim v_1).
\]

This product does not increase the chromatic number.

\begin{lemma}[Sabidussi {\cite[Lemma~2.6]{Sabidussi1957}}]
\label{lem:cartcoloring}
\[
\chi(\G_1 \wedge \G_2) = \max\{\chi(\G_1),\chi(\G_2)\}. 
\]
\end{lemma}

%%%%%%%%%%%%%%%%%%%%%%%%%%%%%%%%%%%%%%%%%%%%%%%%%%%%%%%%%%%%%%%%%%%%%%%%%%%%%%%%
\section{Main Results}
\label{sec:main}

\subsection{Capacity Region via the Confusion Graph}
\label{subsec:def}

We first state a simple generalization of the result by Alon, et al.~\cite[Th.~1.1]{Alon--Hassidim--Lubetzky--Stav--Weinstein2008}.

\begin{proposition}
\label{prop:graphregion1}
A rate tuple $(R_1, \ldots, R_n)$ is achievable for the index coding problem $G$
iff there exists an integer tuple $\tv = (t_1, \ldots, t_n)$ such that
\begin{equation}
\label{eq:graphregion1}
R_j \leq \frac{t_j}{\ceil{\log(\chi(\G_{\tv}(G)))}}, \quad j \in [1:n].
\end{equation}
\end{proposition}

\begin{IEEEproof} \emph{Sufficiency (achievability).} 
For a given tuple $\tv = (t_1, \ldots, t_n)$, consider a coloring of the vertices of the confusion graph 
$\G = \G_{\tv}(G)$ with $\chi(\G)$ colors.
This partitions the vertices of $\G$ into $\chi(\Gc)$ independent sets.
Now by the definition of the confusion graph, no two message tuples in each independent set are confusable
and therefore assigning an index to each independent set yields a valid index code.
The total number of codewords of this index code is $\chi(\G)$, which requires $r = \ceil{\log(\chi(\G))}$ bits to be broadcast. This proves the existence of a $(t_1, \ldots, t_n, \ceil{\log(\chi(\G_\tv(G)))})$ index code.

\emph{Necessity (converse).} 
Consider any $(t_1, \ldots, t_n, r)$ index code, which assigns at most $2^r$ distinct indices to message tuples.
By definition, all the message tuples mapped to an index form an independent set of the
confusion graph $\G = \G_\tv(G)$. Moreover, every message tuple is mapped to some index so that 
these independent sets partition $V(\G)$. 
Thus, $\chi(\G) \le 2^r$, or equivalently, 
$r \ge \ceil{\log(\chi(\G))}$.
Therefore, any achievable $(R_1, \ldots, R_n)$ must satisfy
\[
R_j \le \frac{t_j}{\ceil{\log(\chi(\G_\tv(G)))}}, \quad j \in [1:n],
\]
for some $\tv = (t_1,\ldots,t_n)$.
\end{IEEEproof}

The ceiling operation in~\eqref{eq:graphregion1}, which results from the fact
that the index is communicated in bits, is not essential. By using the 
code that maps $x^n \in \prod_{j=1}^n \{0,1\}^{t_j}$ to $[1:\chi(\G_\tv(G))]$
repeatedly $k$ times, one can easily construct a code that maps
$\xt^n \in \prod_{j=1}^n \{0,1\}^{k t_j}$ to $[1: \chi(\G_\tv(G))]^k$, thus achieving rates
\begin{align}
R_j &= \frac{kt_j}{k\log(\chi(\G_\tv(G)))+1} \notag \\
&\le
\frac{kt_j}{\ceil{k\log(\chi(\G_\tv(G)))}}, \quad j \in [1:n]. \label{eq:ktuple}
\end{align}
Letting $k \to \infty$ in~\eqref{eq:ktuple} establishes the following.

\begin{proposition}
\label{prop:graphregion2}
The capacity region $\Cr$ of the index coding problem $G$ is the closure
of all rate tuples $(R_1, \ldots, R_n)$ such that
\begin{equation}
\label{eq:graphregion2}
R_j \leq \frac{t_j}{\log(\chi(\G_{\tv}(G)))}, \quad j \in [1:n],
\end{equation}
for some $\tv = (t_1, \ldots, t_n)$.
\end{proposition}

We now state a stronger result, in terms of the \emph{fractional} chromatic number,
which will
prove to be useful in establishing structural properties of the capacity region.

\begin{theorem}
\label{thm:graphregion3}
The capacity region $\Cr$ of the index coding problem $G$ is the closure
of all rate tuples $(R_1, \ldots, R_n)$ such that
\begin{equation}
\label{eq:graphregion3}
R_j \leq \frac{t_j}{\log(\chi_f(\G_{\tv}(G)))}, \quad j \in [1:n],
\end{equation}
for some $\tv = (t_1, \ldots, t_n)$.
\end{theorem}

\begin{IEEEproof} The necessity follows by~\eqref{eq:fractional} and Proposition~\ref{prop:graphregion1}.

Let $\e > 0$. For each $\tv = (t_1, \ldots, t_n)$ and the corresponding confusion graph $\G_{\tv}(G)$,
Lemma~\ref{lem:orcoloring1} implies that there exists an integer $k$ such
that
\begin{equation} \label{eq:ktuple2}
\sqrt[k]{\chi(\G_\tv^k(G))} \le \chi_f(\G_{\tv}(G)) + \epsilon.
\end{equation}
It can be also checked that the set of edges of $\G_{\tv}^k(G)$ contains 
the set of edges of $\G_{k \tv}(G)$, which,
when combined with~\eqref{eq:ktuple2}, implies that
$\sqrt[k]{\chi(\G_{k \tv}(G))} \le \chi_f(\G_{\tv}(G)) + \epsilon$,
or equivalently,
\[
\frac{t_j}{\log(\chi_f(\G_{\tv}(G)) + \epsilon)} \le \frac{k t_j}{\log(\chi(\G_{k \tv}(G)))}, \quad j \in [1:n].
\]
Thus, by Proposition~\ref{prop:graphregion2}, if $(R_1,\ldots,R_n)$ satisfies
\[
R_j \le \frac{t_j}{\log(\chi_f(\G_{\tv}(G)) + \epsilon)}, \quad j \in [1:n],
\]
then it must be in the capacity region. Since $\Cr$ is closed, taking $\e \to 0$ completes the proof.
\end{IEEEproof}

%%%%%%%%%%%%%%%%%%%%%%%%%%%%%%%%%%%%%%%%%%%%%%%%%%%%%%%%%%%%%%%%%%%%%%%%%%%%%%%%%%%%%%%%%%%%%%
\subsection{Capacity Region via Confusion Graph Products}
\label{subsec:theorems}

Throughout this subsection, we assume that $G_1$ and $G_2$ are two vertex-induced subgraphs
of $G$ such that $V(G_1) = [1:n_1]$ and $V(G_2) = [n_1+1:n]$ 
partition $V(G) = [1:n]$. We denote the capacity regions of the index
coding problems $G$, $G_1$ and $G_2$ by $\Cr$, $\Cr_1$ and $\Cr_2$, respectively.

\begin{proposition}
\label{prop:unionregion}
If $G$ has no edge between $G_1$ and $G_2$, then
\[
\Cr = \bigcup_{\a \in [0,1]} \bigl\{ (\a \Rv_1 , (1-\a) \Rv_2 ) \suchthat
\Rv_1 \in \Cr_1, \Rv_2 \in \Cr_2 \bigr\}.
\]
\end{proposition}

In other words, the capacity region of $G$ is achieved 
by time division between the optimal coding schemes 
for two disjoint subproblems $G_1$ and $G_2$.

\begin{IEEEproof}
It suffices to show that
\[
\Cr \subseteq \bigcup_{\a \in [0,1]} \bigl\{ (\a \Rv_1 , (1-\a) \Rv_2 ) \suchthat
\Rv_1 \in \Cr_1, \Rv_2 \in \Cr_2 \bigr\}.
\]
Let $x^n = (\xv_1, \xv_2)$ and $z^n = (\zv_1,\zv_2)$ be two message tuples, 
and $\tv = (\tv_1,\tv_2)$ be their common length tuple, where $\xv_i$, $\zv_i$, and $\tv_i$ correspond to the subproblem $G_i$, $i = 1,2$.
By the definition of confusability, $x^n$ and $z^n$ are confusable iff 
they are confusable at some receiver $j \in V(G_1)$
or 
confusable at some receiver $j \in V(G_2)$.
Since there is no edge between $G_1$ and $G_2$, these local confusability
conditions are equivalent to 
the confusability of $\xv_1$ and $\zv_1$ for the subproblem $G_1$ 
and the confusability of $\xv_2$ and $\zv_2$ for the subproblem $G_2$, respectively.
In other words, $x^n$ and $z^n$ are confusable for $G$
iff $\xv_1$ and $\zv_1$ are confusable for $G_1$
or $\xv_2$ and $\zv_2$ are 
confusable for $G_2$.
Thus, $\G_\tv(G) = \G_{\tv_1}(G_1) * \G_{\tv_2}(G_2)$ and by Lemma~\ref{lem:orcoloring} for 
disjunctive product, 
\begin{align*}
&\log(\chi_f(\G_\tv(G))) \\
&= 
\log(\chi_f(\G_{\tv_1}(G_1))) +
\log(\chi_f(\G_{\tv_2}(G_2))) =: l_1 + l_2.
\end{align*}
We now let $\a = l_1 / (l_1+l_2)$ and apply Theorem~\ref{thm:graphregion3}.
Before closure, any rate tuple in $\Cr$ should satisfy
\[
R_j \le 
\frac{t_j}{l_1 + l_2} 
=
\begin{cases}
\a \frac{t_j}{l_1}, & j \in V(G_1),\\
(1-\a) \frac{t_j}{l_2}, & j \in V(G_2).\\
\end{cases}
\]
But again by Theorem~\ref{thm:graphregion3}, $(t_j/l_1: j \in V(G_1)) \in \Cr_1$ and $(t_j/l_2: j \in V(G_2)) \in \Cr_2$, which completes the proof.
\end{IEEEproof}

We now state a stronger version of Proposition~\ref{prop:unionregion}, originally established by
Tahmasbi, Shahrasbi, and Gohari \cite{Tahmasbi--Shahrasbi--Gohari2014b}; see also
\cite[Th.~8]{Bar-Yossef--Birk--Jayram--Kol2011} for a related but much weaker statement.

\begin{theorem}[{Tahmasbi, Shahrasbi, and Gohari~\cite[Th.~2]{Tahmasbi--Shahrasbi--Gohari2014b}}]
\label{thm:USCS}
If $G$ has no edge from $G_2$ to $G_1$, then
\[
\Cr = \bigcup_{\a \in [0,1]} \bigl\{ (\a \Rv_1 , (1-\a) \Rv_2 ) \suchthat
\Rv_1 \in \Cr_1, \Rv_2 \in \Cr_2 \bigr\}.
\]
\end{theorem}

Once again the capacity region is achieved by time division. Moreover, in light of Proposition~\ref{prop:unionregion}
and the Farkas lemma~\cite[Th.~2.2]{Achim--Kern1992} (that is, each edge in a directed graph either lies on a directed cycle or belongs to a directed cut but not both), Theorem~\ref{thm:USCS} implies that removing edges of $G$
that do not lie on a directed cycle does not change the capacity region.

\begin{IEEEproof}
Assume without loss of generality that there exists an edge from every node in $G_1$ to every node in $G_2$.
Now, since every node in $G_2$ has every node (message) in $G_1$ as side information
and no node in $G_1$ has any node in $G_2$ as side information,
$x^n$ and $z^n$ are confusable for $G$ iff $\xv_1$ and $\zv_1$ are confusable for $G_1$, or 
$\xv_1 = \zv_1$ and
$\xv_2$ and $\zv_2$ are 
confusable for $G_2$ (recall the notation in the proof of Proposition~\ref{prop:unionregion}).
Thus, $\G_\tv(G) = \G_{\tv_1}(G_1) \lexprod \G_{\tv_2}(G_2)$ and by Lemma~\ref{lem:lexcoloring} for 
lexicographic product, 
\begin{align*}
&\log(\chi_f(\G_\tv(G))) \\
&\qquad= 
\log(\chi_f(\G_{\tv_1}(G_1))) +
\log(\chi_f(\G_{\tv_2}(G_2))).
\end{align*}
The rest of the proof follows the identical steps to that of Proposition~\ref{prop:unionregion}.
\end{IEEEproof}

The only difference between Propposition \ref{prop:unionregion} and Theorem \ref{thm:USCS} lies with which product of confusion subgraphs needs to be taken---the disjunctive product for two separate subproblems, while the lexicographic product for two subproblems dependent in only one direction. Note that the main tool we use
from fractional graph theory
(cf.~Lemmas~\ref{lem:orcoloring} and \ref{lem:lexcoloring})
is
\begin{equation} \label{eq:key-ineq}
\chi_f(\G_\tv(G)) \ge \chi_f(\G_{\tv_1}(G_1)) \chi_f(\G_{\tv_2}(G_2)).
\end{equation}
This implies that as long as an index coding problem can be partitioned into two subproblems and the corresponding (nonstandard) graph product of the confusion subgraphs satisfies \eqref{eq:key-ineq}, the capacity region has the same form as in Proposition~\ref{prop:unionregion} and Theorem~\ref{thm:USCS}. Also note that 
with the (integral) chromatic number, an inequality like \eqref{eq:key-ineq} holds in the opposite direction 
(cf.~\eqref{eq:orcoloring2} and \eqref{eq:lexcoloring2}). This shows the major advantage of Theorem~\ref{thm:graphregion3} over Proposition~\ref{prop:graphregion2}.

Next, we consider index coding problems 
with side information graphs that contain a complete bipartite graph as an edge-induced subgraph.

\begin{theorem}
\label{thm:intersectionregion}
If there are edges from every node in $G_1$ to every node in $G_2$ and vice versa,
then
\[
\Cr = \bigl\{(\Rv_1,\Rv_2) \suchthat \Rv_1 \in \Cr_1, \Rv_2 \in \Cr_2 \bigr\}.
\]
\end{theorem}

In other words, the capacity region of $G$ is achieved 
by simultaneously using the optimal coding schemes 
for two disjoint subproblems $G_1$ and $G_2$.

\begin{IEEEproof} 
Since every node in $G_1$ has every message in $G_2$ as side information and every node in $G_2$ has every message in $G_1$ as side information, $x^n$ and $z^n$ are confusable for $G$ iff $\xv_1 = \zv_1$ and $\xv_2$ and $\zv_2$ are confusable for $G_2$, or $\xv_2 = \zv_2$ and $\xv_1$ and $\zv_1$ are confusable for $G_1$.
Thus,
$\G_\tv(G) = \G_{\tv_1}(G_1) \wedge \G_{\tv_2}(G_2)$ and by Lemma \ref{lem:cartcoloring} for cartesian product, 
\begin{equation}
\label{eq:cartprod}
\chi(\G_{\tv}(G)) = \max\{\chi(\G_{\tv_1}(G_1)),\chi(\G_{\tv_2}(G_2))\}.
\end{equation}
By Proposition \ref{prop:graphregion2}, before closure, any rate tuple in $\Cr$ should satisfy
\[
R_j \le 
\frac{t_j}{\log(\chi(\G_{\tv}(G)))}, \quad j \in [1:n],
\]
for some $\tv = (t_1, \ldots, t_n)$.
Combining this with (\ref{eq:cartprod}), we have for $i = 1,2$
\[
R_j \le 
\frac{t_j}{\log(\chi(\G_{\tv_i}(G_i))} =: \frac{t_j}{l_i}, \quad j \in V(G_i).
\]
By applying Proposition~\ref{prop:graphregion2} once again, $(t_j/l_1: j \in V(G_1)) \in \Cr_1$ and $(t_j/l_2: j \in V(G_2)) \in \Cr_2$, which completes the proof. 
\end{IEEEproof}

\subsection{Capacity Region via Degraded Side Information Sets}

Here we consider the index coding problem $G$ with side information sets $A_1, \ldots, A_n$.

\begin{proposition}
\label{prop:ext-side-info}
If $A_i \subseteq A_j$, then removing $i$ from $A_j$ does 
not decrease the capacity region.
\end{proposition}

The proof is intuitively clear. Given any index code, receiver~$j$ can first recover $x_i$ using $A_i$
and then uses $x_i$ along with $x(A_j \setminus \{i\})$ to recover $x_j$.
Here is an alternative proof based on the notion of confusion graph.

\begin{IEEEproof}
Assume that there exist $x^n$ and $z^n$ confusable for the new problem $G'$,
but not for the original problem $G$. Then, they must be confusable at receiver $j$ for $G'$
(i.e., $x_j \ne z_j$ and $x(A_j \setminus \{i\}) = z(A_j \setminus \{i\})$).
Now if $x_i = z_i$, then it contradicts the assumption that $x^n$ and $z^n$ are not confusable (at receiver $j$) for $G$. Alternatively, if $x_i \ne z_i$, then since $A_i \subseteq A_j$ and hence $x(A_i) = z(A_i)$, it again contradicts the assumption that $x^n$ and $z^n$ are not confusable (at receiver $i$) for $G$. Therefore, the confusion graphs must be the same and, by Theorem~\ref{thm:graphregion3}, so must be the capacity regions.
\end{IEEEproof}

%%%%%%%%%%%%%%%%%%%%%%%%%%%%%%%%%%%%%%%%%%%%%%%%%%%%%%%%%%%%%%%%%%%%%%%%%%%
\bibliographystyle{IEEEtran}
\bibliography{nit} 

\end{document}